# Sparse Sampling for Fast Quasiparticle Interference Mapping

Jens Oppliger and Fabian Donat Natterer*

Department of Physics, University of Zurich, CH-8057 Zurich, Switzerland
*To whom all correspondence shall be addressed: fabian.natterer@uzh.ch

**Scanning tunneling microscopy (STM) is a notoriously slow technique; Data-recording is serial which renders complex measurement tasks, such as quasiparticle interference (QPI) mapping[1–12], impractical. However, QPI would provide insight into band-structure details of quantum materials which can be inaccessible to angle-resolved photoemission spectroscopy. Here we use compressed sensing (CS)[13,14] to fundamentally speed-up QPI mapping. We reliably recover the QPI information from a fraction of the usual local density of state measurements. The requirement of CS is naturally fulfilled for QPI, since CS relies on sparsity in a vector domain, here given by few nonzero coefficients in Fourier space. We exemplify CS on a simulated Cu(111) surface using random sampling of constant and varying probability density. We further simplify the motion of the STM tip through an open traveling salesman's problem for greater efficiency[15]. We expect that the implications of our CS approach will be transformative for the exploration of 2D quantum materials.**

The saying *time is of the essence* also applies to the study of condensed matter systems, especially when experimental discovery trails the theoretical prediction of novel quantum materials[16–18]. This new paradigm of *theory leading experiment* has made an important showing for topological insulators[19] and most recently in twisted van-der-Waals materials[20]. The ever-growing list of predicted candidate systems needs a matching effort to verify their properties.

Recent work has made important strides at the higher level of feature analysis using machine learning[6] while the preceding data recording oftentimes remains on the sideline. Angle-resolved photoemission spectroscopy (ARPES)[21–23] has hereby served as the prime tool for band-structure investigations, routinely providing valuable insight into the electronic structure of candidate materials. However, the study of advanced materials also requires the preparation of conditions in which ARPES is inhibited. These limitations are particularly concerning in the view of micron sized samples, field-effect devices, magnetic fields, cryogenic temperatures, and hole doped systems; all of which could trigger quantum phase transitions and that are natural control knobs in the exploration of quantum materials. Fortunately, a scanning tunneling microscope excels under these conditions. An STM is naturally surface sensitive, operates down to milli-Kelvin temperatures[24,25] at microelectronvolt energy resolution, in high magnetic fields, and on gate tunable devices. The STM obtains in-plane momentum sensitivity through the elastic scattering of electrons at surface discontinuities (such as atomic steps and point defects) that modulate the local density of states (LDOS)[1,26,27] which resemble standing wave patterns. These LDOS modulations provide access to the scattering space ($q$-space)[3,4] via the Fourier transform, from which momentum space information can be inferred ($k$-space). The band-structure mapping from LDOS modulations is known as Fourier-transform or quasiparticle interference (QPI) STM. Although QPI-STM appears viable for band structure determination, it presently is also excruciatingly slow, rendering the comprehensive study of a larger parameter space (temperature, magnetic field, doping) inconceivable.

The throughput of a QPI investigation is presently constrained by the following considerations: In order to distinguish nuances in reciprocal space, the finest momentum space resolution of $q_{min} = 2\pi/L$, where $L$ stands for the physical size of the LDOS map, as well as a large number $n$ of its increments, is desired. However, the total number $N$ of point-spectra that are used to map the $LDOS(x, y, E)$ grows quadratically with the grid size $N = n \times n$, which quickly amounts to hundreds of thousands of spectra and daylong measurements[12], during which the experiment hinges at the mercy of a notoriously unpredictable STM tip.

Despite these measurement efforts, the resulting QPI patterns contain only a few nonzero coefficients as can be seen in Figure 1 and previous work[1–11], suggesting a high degree of redundancy in conventional QPI mapping. Pointing out the undeniable sparsity in the QPI pattern reveals an important connection to the key requirement of compressive sensing (CS)[13,14]. Compressive sensing uses sparsity of information in some vector representation and incoherent sampling to enable a new data recording paradigm that requires radically fewer measurements than conventional sampling in the Nyquist regime. A suitable analogy for the present QPI case would be the sparse sampling of a sine wave in time-domain (mimicking spatial LDOS modulations) and its CS recovery of the only two coefficients in Fourier space (equivalent to the sparse QPI pattern).

In the following, we exploit the sparsity of the QPI pattern to fundamentally speed up QPI mapping using a compressed sensing approach that we combine with an efficient STM tip-routing between randomly located LDOS measurements. Figure 1 exemplifies the difference between conventional QPI mapping and our compressed sensing method on a deliberately small $N = 64 \times 64$ grid for better visualization. The simulated copper surface-state shows a standing wave pattern from an individual point scatterer that modulates the $LDOS(x, y, E)$ and that we further perturb by white-noise. From the fully sampled LDOS (conventional QPI), the Fourier transform produces the QPI pattern showing a ring of radius $q = 2k$ and six Bragg spots corresponding to the Cu(111) lattice. In perfect agreement, we recover the same QPI pattern from the sparsely sampled LDOS seen in Figure 1 using only 20% of random LDOS measurements. We obtain the original LDOS by a Fourier transform of the recovered QPI in Figure 1c, which highlights the excellent CS reconstruction and complimentary noise rejection. Since the locations at which we randomly measure the LDOS have to visited by the STM tip, we interpret the path of the STM tip between those points as an open traveling salesman's problem (TSP) to save time and for which we find a near-optimal solution using a genetic algorithm[15]. The TSP further introduces a useful time-correlation in the sampling data that could be used to correct for drift. Overall, this simple example reveals the fundamental time-savings of our CS method at a fivefold advantage compared to conventional QPI mapping, while the TSP optimized STM tip-path reduces travel distance of the tip by another ~63% with respect to the fully sampled case.

We next unleash the working principle of our compressed sensing method onto a situation which was previously impossible in conventional QPI mapping to illustrate the game-changing advantage for quantum materials exploration. Figure 2 shows a large image on a grid of size $N = 1024 \times 1024$ which we sparsely sample using only 2, 6, and 18% of the LDOS data. The compressed sensing recovery using random-point-sampling achieves QPI reconstruction down to a sparse sampling of only 5% (Figure 2c and f). Assuming 1 second per LDOS measurement as a time reference, we now contrast conventional QPI and our CS mapping: The former would accordingly take 291 hours with an additional ~30 hours to solely move the measurement locations assuming 0.1 s per hop, by far exceeding the hold time of state-of-the-art STM facilities (see comparison chart in Ref.[25]). In contrast, our CS method would take a mere ~14.5 hours for sampling 5%, and 2 to 6 hours for tip-travel, ideally suited for overnight operation or for the systematic investigation of a large parameter space, for instance via magnetic field and temperature dependence.

We now explore a sparse sampling variant, referred to as 'informed sampling (IS)', that requires even fewer samples for QPI recovery than the just established random-point-sampling (Figure 2b). Upon reexamination of the LDOS in Figure 1a we note that the most dominant modulations are in the immediate vicinity of the scattering sites, decaying rapidly into the noise at increasing distance. Since these scattering sites are well-known from explorative topographies that typically precede QPI measurements, we specifically mark their $(x, y)$ locations around which we enforce a denser but still random sampling mask indicated by the green circles in Figure 2. The comparison with random-point-sampling in Figure 2c,d shows that IS achieves QPI recovery even at 2% subsampling when random-point-sampling has already failed. Our IS would provide a solution to difficult QPI cases, in presence of too few scatterers, or when there are time restraints. Hypothetically, the present IS example would take only ~6 hours, equivalent to a 50-fold faster QPI mapping.

We finally utilize informed sampling to efficiently deal with an inverted situation in which we intend to avoid regions instead of sampling them more densely. This could be required to exclude the influence of step-edges[12] or mask point impurities[7] that provoke tip-instabilities or lead to diffuse LDOS modulations. Since the LDOS data of those regions would be removed in conventional QPI after their measurement, our informed sampling saves additional resources by not measuring them in the first place. Figure 3 shows how we reliably achieve QPI recovery for this sampling mode and how we implement our traveling salesman's interpretation of the tip-path.

In conclusion, we have introduced an efficient sparse sampling method for quasiparticle interference mapping with the STM. Our method requires no additional hardware and can readily enhance the operation of existing STM facilities. The fundamental time-savings of our method renders previously inconceivable QPI experiments possible; our method enables the exploration of an otherwise inaccessibly large parameter space for the characterization of advanced quantum materials. The suggested informed sampling variant provides further enhancement by either concentrating the random sampling around strong scatterers or by deliberately avoiding regions that would be discarded in conventional sampling after the fact. Our traveling salesman's approach for optimizing the path length of the STM tip significantly reduces additional measurement overhead.

**Methods**

In order to use our compressed sensing method for quasiparticle interference mapping we proceed as follows: We first generate a sampling matrix on a grid of size $N = n \times n$. Depending on our sampling variant (random-point-sampling or informed-sampling), we exclude regions or favor entries that we randomly set to "1" for a fraction $p$ of the entries, while the other $1 - p$ remain "0". The "1" denote the random locations at which we measure the LDOS during our sparse mapping. We next compute a near-optimal solution to the open traveling salesman's problem using a genetic algorithm[15]. Since the traveling salesman's problem is NP-hard, the calculation of an optimal path for large sampling matrices may become cumbersome but a catalogue of sampling masks could be calculated long before the actual QPI experiment for a range of $p$ values and grid sizes $N$. The coordinates would next be fed in their optimal order to the STM controller for the sparse LDOS mapping, generating a measurement vector $b$ of size $pN \times 1$. From these sparsely sampled LDOS measurements, we next compute the QPI pattern using the Matlab based compressed sensing solver SPGL1[28,29] and its basis pursuit denoising variant, which aims at minimizing $\|x\|_1$ subject to $\|Ax - b\|_2 \leq \sigma$. Here, $x$ represents the $s$-sparse $N \times 1$ QPI pattern, $A$ is the $pN \times N$ measurement matrix connecting real and momentum space, and $\sigma$ is an adjustable noise tolerance threshold. Note that we cast $n \times n$ dimensional problem into a one-dimensional $N \times 1$ description, which we bring back to 2D after recovery and for illustration. The CS recovery directly yields the QPI pattern that we Fourier transform to obtain the LDOS modulations in real space. For explorative closed-loop LDOS measurements at a single energy, a TSP or sparse line-hopping[30] could provide provide a fast overview of the QPI information using CS.

## Acknowledgements

F.D.N. greatly appreciates support from the Swiss National Science Foundation under project number PZ00P2_176866. We thank Thomas Greber, Guiseppe Genovese and Vidya Madhavan for fruitful discussions.

## Author information

F.D.N. conceived and supervised the project. J.O. carried out the compressed sensing simulations. F.D.N. wrote the manuscript with support of J.O.. All authors discussed the results.

## Competing interests

The authors declare no competing interests.

**Figure Captions**

**Figure 1 | Demonstrating the Sparse Sampling for Quasiparticle Interference mapping. a,** In conventional QPI mapping, the scanning tunneling microscope (STM) records the local density of state (LDOS) modulations on every surface point, here on a simulated 64 × 64 grid of a simulated Cu(111) surface with a gaussian noise of 0.2 of the LDOS standard deviation added. Using a sparse-sampling method, we only sample a small fraction of the full LDOS (here 20%) by efficiently moving the tip according to a traveling salesman between the distributed measurement locations. **b,** The QPI pattern is the Fourier transform of the LDOS modulations as shown for the fully sampled LDOS of the conventional method (top) and from our 20% sparsely sampled LDOS. The wavevector of the surface state as well as the Bragg peaks of the Cu(111) lattice are identical in the original and recovered QPI pattern. **c,** The QPI pattern allows the illustration of the original LDOS modulations via inverse Fourier transform, showing the excellent recovery and noise rejection for the sparsely sampled mapping.

**Figure 2 | Large scale random and informed sampling.** Local density of states modulations created by 10 simulated scattering sites are sampled for 2% of the LDOS using **a** random and **b** informed sampling. For the latter, the sampling is denser around point-scatterers, marked by rings in (a,b). **c-d,** recovered QPI patterns from sparse sampling at different fractions of LDOS measurements with sparsity and number of recovered

coefficients indicated. Informed Sampling better recovers QPI information until about 20% LDOS measurements, as seen in the plots in e and f. **e**, optimal radius (HWHM) for a Lorentzian probability distribution used in the informed sampling method. **f**, Number of recovered QPI coefficients for random and informed sampling also shows the cross-over to random-sampling around 20%.

**Figure 3 | Informed sampling masking regions excluded from QPI mapping. a,** LDOS overlaid with near-optimal TSP tip-path indicated by the black line showing starting (A) and endpoints ($\Omega$). Sparsely sampled LDOS at 8% subsampling on a $128 \times 128$ grid, with excluded regions marked by hue circles. **b**, Comparison of QPI patterns obtained from a fully (top) and sparsely sampled (bottom) LDOS mapping. Our informed sampling recovery reliably operates also without sampling the regions masked in **a**.

**Figure 1 |**

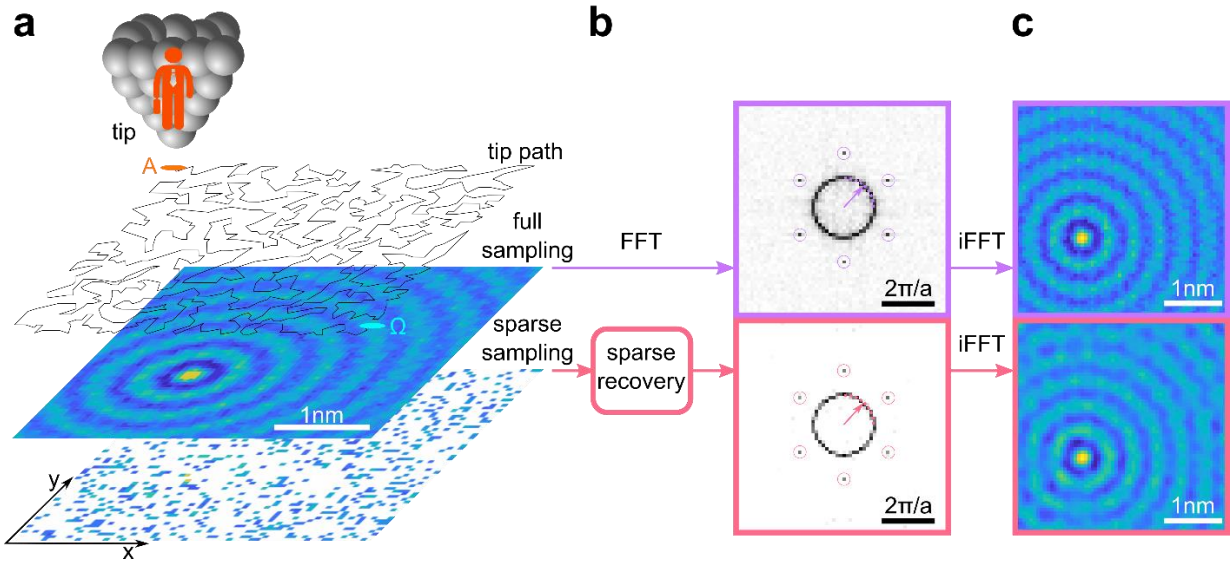

*Figure 1*

**Figure 2 |**

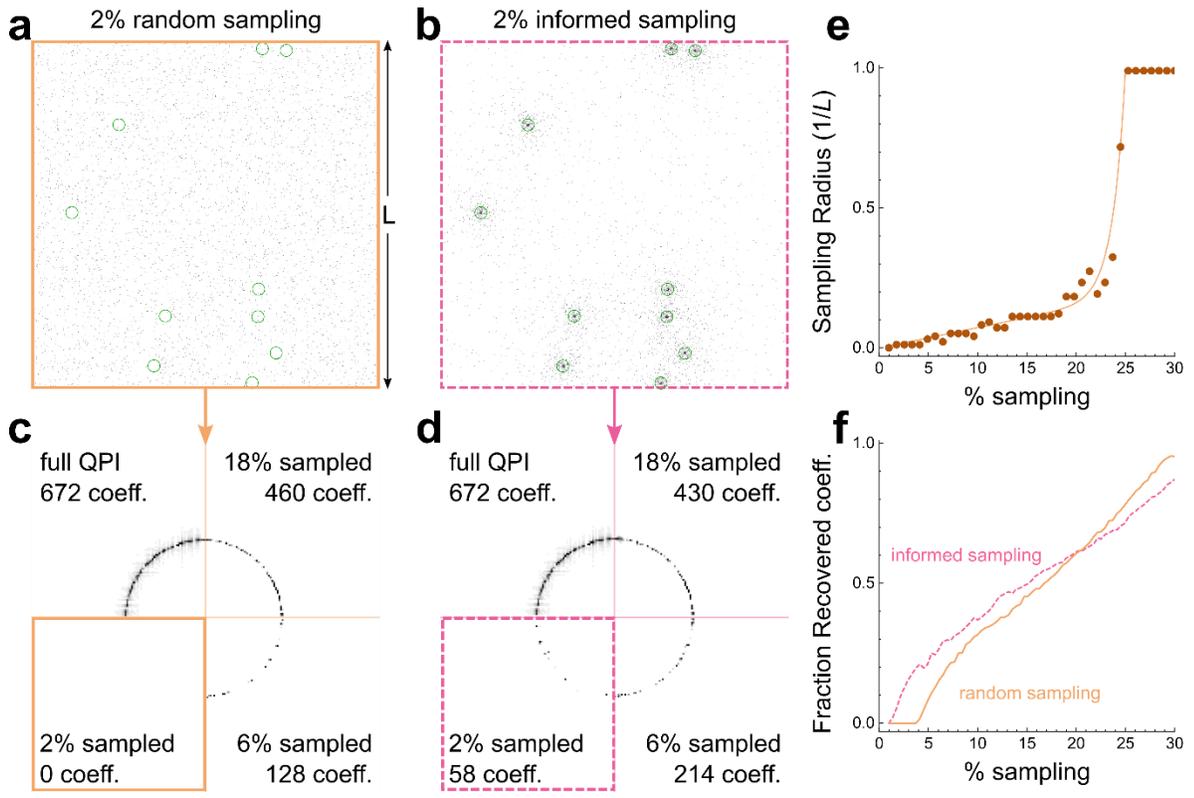

*Figure 2*

**Figure 3 |**

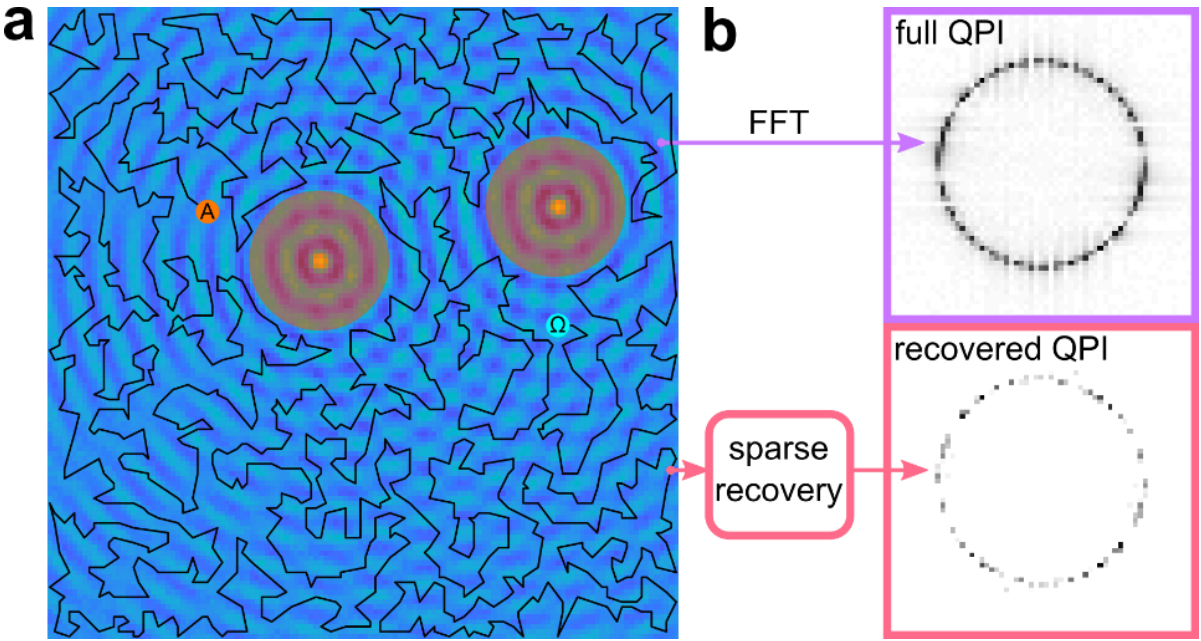

*Figure 3*